\documentclass[aps,pre,twocolumn,amssymb,amsfonts,floatfix,showpacs,superscriptaddress]{revtex4}

\usepackage{graphicx}
\usepackage{dcolumn}
\usepackage{bm}
\usepackage{mathbbol}
\usepackage{graphics,psfrag}
\usepackage{amsmath,amssymb,amsthm}
\usepackage{times}
\usepackage{bm}
\usepackage[dvips]{color}

\begin{document}

\title{Time Fluctuations in Isolated Quantum Systems of Interacting Particles}

\author{Pablo R. Zangara}
\affiliation{Instituto de F\'{i}sica Enrique Gaviola, CONICET-UNC and Facultad
de Matem\'{a}tica, Astronom\'{i}a y F\'{i}sica, Universidad Nacional
de C\'{o}rdoba, 5000, C\'{o}rdoba, Argentina}
\author{Axel D. Dente}
\affiliation{Instituto de F\'{i}sica Enrique Gaviola, CONICET-UNC and Facultad
de Matem\'{a}tica, Astronom\'{i}a y F\'{i}sica, Universidad Nacional
de C\'{o}rdoba, 5000, C\'{o}rdoba, Argentina}
\author{E. J. Torres-Herrera}
\affiliation{Department of Physics, Yeshiva University, New York, NY 10016, USA}
\author{Horacio M. Pastawski}
\affiliation{Instituto de F\'{i}sica Enrique Gaviola, CONICET-UNC and Facultad
de Matem\'{a}tica, Astronom\'{i}a y F\'{i}sica, Universidad Nacional
de C\'{o}rdoba, 5000, C\'{o}rdoba, Argentina}
\author{An\'{i}bal Iucci}
\affiliation{Instituto de F\'{i}sica La Plata - CONICET and Departamento de F\'{i}sica, Universidad Nacional de La Plata, cc 67, 1900 La Plata, Argentina.}
\author{Lea F. Santos}
\affiliation{Department of Physics, Yeshiva University, New York, New York 10016, USA}

\begin{abstract}
Numerically, we study  the time fluctuations of few-body observables after relaxation in isolated dynamical quantum systems of interacting particles. Our results suggest that they decay exponentially with system size in both regimes, integrable and chaotic. The integrable systems considered are solvable with the Bethe ansatz and have a highly nondegenerate spectrum. This is in contrast with integrable Hamiltonians mappable to noninteracting ones. We show that the coefficient of the exponential decay depends on the level of delocalization of the initial state with respect to the energy shell.
\end{abstract}

\maketitle

\section{Introduction}

The nonequilibrium dynamics of quantum many-body systems is a challenging  and little understood subject of modern physics.  Step by step, numerical, analytical, and experimental studies have been trying to put the pieces of the puzzle together by identifying properties and behavior common to different quantum systems. Our main goal in this paper is the search for a general picture of the behavior of few-body observables in isolated quantum many-body systems after equilibration. In particular, we investigate whether their time fluctuations depend on regime, initial states, and observables. 

Equilibration in isolated quantum systems can happen in a probabilistic sense. It requires that: (i) the time fluctuations of the observables, after the transients have died out, become very small, implying proximity to the stationary state for the vast majority of time, and (ii) the fluctuations decrease with system size, vanishing in the thermodynamic limit. Based on semiclassical arguments and on fully chaotic systems~\cite{Feingold1986,Deutsch1991,Prosen1994,SrednickiARXIV,Srednicki1996,Srednicki1999}, it has been shown that the mean squared amplitude of temporal fluctuations after relaxation decrease exponentially with system size. This derivation is independent of the details of the initial state, which is assumed to be an arbitrary pure state~\cite{Srednicki1996,Srednicki1999}. However, the underlying association with full random matrices overrides some of the facets of finite real systems, which are of relevance to experiments. Real systems have few-body and usually short-range interactions, whereas full random matrices imply many-body long-range interactions~\cite{Brody1981,ZelevinskyRep1996,Kota2001}. In real systems, the density of states is Gaussian~\cite{Brody1981}, so chaotic eigenstates, where the probability amplitudes of the basis vectors are many, small, and uncorrelated, are restricted to the middle of the spectrum. More recent studies for the bounds of the time fluctuations relaxed the condition on full random matrices and relied on Hamiltonians without too many degeneracies of eigenvalues and energy gaps and on initial states made of large superpositions of energy eigenstates~\cite{Reimann2008,Short2011,Short2012,Reimann2012}. The fluctuations were again shown to scale exponentially with system size. Yet, in the particular case of an integrable Hamiltonian quadratic in the canonical Fermi operators or mapped onto one, where the nonresonant conditions are not satisfied, it was shown analytically~\cite{Venuti2013} and numerically~\cite{Cassidy2011,Gramsch2012,HeARXIVdisorder} that the time fluctuations of one-body or quadratic observables scale as $1/\sqrt{L}$, $L$ being the system size. 

These findings motivate the questions: How do the time fluctuations scale with $L$ in the case of integrable systems that cannot be mapped to free particles? How about chaotic systems where the energy of the initial state is far from the middle of the spectrum? We explore these questions with one-dimensional spin-1/2 models in both integrable and chaotic domains. Our results indicate that  the answer for the two questions is, once again, exponential scaling. Integrable models not mappable onto free particle systems are significantly less degenerate than noninteracting ones. As for initial states close to the edges of the spectrum and, therefore, not substantially delocalized in the energy representation, the coefficient of the exponential decay becomes small, but exponential fittings are still better than power law. It is only in the case of the noninteracting spin-1/2 model that the power-law fitting is superior, provided the initial state is not thermal. 

To evaluate the level of delocalization of the initial state, we employ the concept of the energy shell as established in many-body quantum chaos~\cite{Casati1993,Casati1996}. In this field, the total Hamiltonian of the system is often separated in an unperturbed part, which describes the noninteracting particles (or quasiparticles), and a perturbation, which represents the inter-(quasi)particle interactions and may drive the system into the chaotic domain. The Hamiltonian matrix is then written in the basis corresponding to the unperturbed vectors (the mean-field basis). The distribution in energy of the components $C_{\alpha}^j$ of the mean-field basis vectors $|j \rangle= \sum_{\alpha} C_{\alpha}^j |\alpha \rangle$, $|\alpha \rangle$ being the eigenstates of the total Hamiltonian, is known as the strength function or local density of states~\cite{Flambaum2000}. The energy shell corresponds to the maximal strength function obtained in the limit of very strong interactions. The energy shell has a Gaussian shape and a dual role: It determines the maximum possible spreading of the unperturbed states in the energy representation, as well as the maximum level of delocalization of the eigenstates in the mean-field basis. In real systems with few-body finite-range interactions, the states become more delocalized as the perturbation increases, but they do not get totally extended, as in full random matrices. Chaotic states are then defined as states that fill the energy shell ergodically, so that their components can be seen as random variables following a Gaussian distribution~\cite{Casati1993,Casati1996,Santos2012PRL,Santos2012PRE}. 

We borrow the ideas above and apply them to the context of nonequilibrium dynamics. The total Hamiltonian dictating the dynamics of the system is written in a basis that incorporates the initial state as one of its vectors. The width of the energy distribution of the initial state corresponds to the width of what we call, here, the energy shell. The lifetime of the initial state depends on how large this width is and on the filling of the shell. As we show, when the width of the energy shell is small compared to the width of the density of states and when it is not well filled, which happens for initial states close to the edges of the spectrum, the relaxation process can become very slow. This scenario is further aggravated by integrable Hamiltonians, the presence of symmetries, and the observable studied.

This paper is organized as follows. In Sec.~II, we describe the spin-1/2 model, the initial states, and the observables considered, as well as the numerical method employed for the time evolution. Section III examines the spectrum and the level of delocalization of the initial states with respect to the energy shell. The scaling analysis of the time fluctuations with system size are presented in Sec.~IV, and results for the relaxation process are discussed in Sec.~V. Concluding remarks are made in Sec.VI.

\section{System and Quantities Studied} 

We consider a one-dimensional lattice of interacting spins 1/2 with open boundaries and an even number $L$ of sites. The Hamiltonian contains nearest-neighbor (NN) and possibly also next-nearest-neighbor (NNN) couplings,
\begin{eqnarray}
&& \hat{H} = \hat{H}_{NN} + \lambda \hat{H}_{NNN} ,
\label{ham} \\
&& \hat{H}_{NN} = \sum_{n=1}^{L-1} J \left(\hat{S}_n^x \hat{S}_{n+1}^x + \hat{S}_n^y \hat{S}_{n+1}^y +\Delta \hat{S}_n^z \hat{S}_{n+1}^z \right) ,
\nonumber \\
&& \hat{H}_{NNN} = \sum_{n=1}^{L-2} J \left(\hat{S}_n^x \hat{S}_{n+2}^x + \hat{S}_n^y \hat{S}_{n+2}^y +\Delta \hat{S}_n^z \hat{S}_{n+2}^z \right) .
\nonumber 
\end{eqnarray}
Above,  $\hbar=1$ \cite{footnote} and $\hat{S}^{x,y,z}_n$ are the spin operators on site $n$. The coupling strength $J$ determines the energy scale and is set to 1, the anisotropy $\Delta$ and the ratio $\lambda$ between NNN and NN exchanges are positive. The flip-flop term $\hat{S}_n^x \hat{S}_{n+1}^x + \hat{S}_n^y \hat{S}_{n+1}^y$ $(\hat{S}_n^x \hat{S}_{n+2}^x + \hat{S}_n^y \hat{S}_{n+2}^y)$ moves the excitations through the chain and $\hat{S}_n^z \hat{S}_{n+1}^z (\hat{S}_n^z \hat{S}_{n+2}^z)$ corresponds to the Ising interaction between NN (NNN) spins. The Hamiltonian conserves total spin in the $z$ direction, $[\hat{H},{\cal \hat{S}}^z]=0$, where ${\cal \hat{S}}^z = \sum_{n=1}^L \hat{S}_n^z$. Other symmetries include parity, invariance under a global $\pi$ rotation around the $x$ axis when  ${\cal \hat{S}}^z=0$, and conservation of total spin  ${\cal \hat{S}}^2=(\sum_{n=1}^L \vec{S}_n)^2$ when $\Delta=1$. The model is integrable and is solved with the Bethe ansatz when $\lambda=0$ \cite{Bethe1931}, and it undergoes a crossover to the chaotic regime as $\lambda $ increases~\cite{Gubin2012,Santos2012PRE}. 

The properties of the spin-1/2 chain (\ref{ham}) depend on the values of the parameters. Accordingly, a nomenclature was developed for different special points. For $\lambda=0$, the noninteracting model ($\Delta=0$) is usually referred to as the XX model, whereas for $\Delta \neq 0$, it is known as the XXZ model \cite{Mikeska}. At the isotropic point $\Delta=1$, the XXZ model is sometimes called XXX.  The value of $\Delta$ determines whether the flip-flop term or the Ising interaction is dominant. When $|\Delta| >1$, an energy gap between the lowest eigenvalues and the ground state appears and the system is said to be in the gapped phase. $\Delta=1$ is the critical point separating the gapped from the gapless phase. As the value of $\Delta $ decreases from 1 to 0, bound states of quasiparticles progressively dissolve into elementary excitations, until the free fermion limit is reached. In this process, a quantitative change in the spectrum occurs at the midpoint $\Delta =1/2$ where the system develops additional nontrivial symmetries \cite{Franchini,Sato,Takahashi}. 

We investigate the dynamics of the system for the following choices of parameters: 

$\bullet $ Integrable isotropic NN Hamiltonian, $\hat{H}_{\Delta=1,\lambda=0}$.

$\bullet $ Integrable anisotropic NN Hamiltonian, $\hat{H}_{\Delta=0.5,\lambda=0}$.

$\bullet $ Weakly chaotic isotropic Hamiltonian, $\hat{H}_{\Delta=1,\lambda=0.4}$.

$\bullet $ Strongly chaotic isotropic Hamiltonian, $\hat{H}_{\Delta=1,\lambda=1}$.

$\bullet $ Strongly chaotic anisotropic Hamiltonian, $\hat{H}_{\Delta=0.5,\lambda=1}$.

The gapped NN Hamiltonian $\hat{H}_{\Delta=1.5,\lambda=0}$ and the noninteracting case $\hat{H}_{\Delta=0,\lambda=0}$ are also discussed on certain occasions, but are not the focus of this paper.

\begin{figure}[htb]
\centering
\includegraphics*[width=0.45\textwidth]{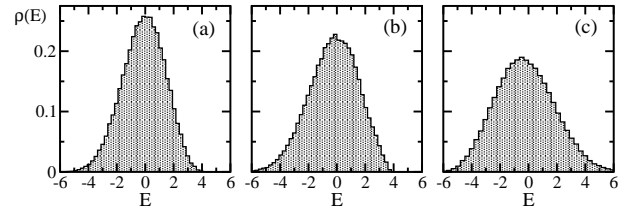}
\caption{Density of states, $L=16$, ${\cal S}^z=0$. (a) $\Delta=0.5, \lambda=0$; (b) $\Delta=1, \lambda=0$; and (c) $\Delta=0.5, \lambda=1$.}
\label{fig:DOS}	
\end{figure}

Independent of the regime of our system, the density of states has a Gaussian shape, as seen in Fig.~\ref{fig:DOS}. This is typical of systems with few-body interactions and is in clear contrast with the semicircular density of states obtained with full random matrices~\cite{HaakeBook,Guhr1998,ReichlBook}.  The Gaussian shape reflects the reduced numbers of energy levels available in the edges of the spectrum. Delocalized states are, therefore, not to be found too far away from the middle of the spectrum, even when the system is chaotic. Notice also that the distributions are not exactly symmetric when $\Delta\neq0$. The tail gets more extended to low energies when $\lambda=0$ and $\Delta $ increases, whereas it goes further to the right when $\lambda > \Delta$.

The width $\omega $ and the average energy $\langle E \rangle $ obtained from a Gaussian fit for the Hamiltonians studied here are shown in Table~\ref{table:DOS}. The density of states obviously gets broader with the anisotropy and the inclusion of NNN couplings. Its center is also displaced from zero as the Ising interaction increases.
\begin{table}[h]
\caption{Width and center of the Gaussian fit for the density of states; $L=16$; ${\cal S}^z=0$.}
\begin{center}
\begin{tabular}{|l|c|c|}
\hline 
  &  \hspace{0.3 cm} $\omega $ \hspace{0.3 cm} &  \hspace{0.2 cm} $\langle E \rangle$  \hspace{0.2 cm}  \\
  \hline
$\hat{H}_{\Delta=0,\lambda=0}$ &  1.444  &    0.000   \\
\hline
$\hat{H}_{\Delta=0.5,\lambda=0}$ &  1.532  &  -0.039   \\
\hline
$\hat{H}_{\Delta=1,\lambda=0}$ &  1.761  &   -0.119    \\
\hline
$\hat{H}_{\Delta=1.5,\lambda=0}$ &  2.078  &  -0.234     \\
\hline
$\hat{H}_{\Delta=1,\lambda=0.4}$ &  1.868  & -0.368      \\
\hline
$\hat{H}_{\Delta=1,\lambda=1}$ &  2.399 &   -0.571    \\
\hline
$\hat{H}_{\Delta=0.5,\lambda=1}$ &  2.108  &  -0.356     \\
\hline
\end{tabular}
\end{center}
\label{table:DOS}
\end{table}

\subsection{Initial states}

The Hamiltonian matrix is written in a basis in which each site has a spin either pointing up or pointing down  in the $z$ direction. These vectors correspond to the eigenstates of the Ising part of the Hamiltonian. We refer to it as the site basis. The system is prepared in an initial state $|\Psi(0)\rangle \equiv |{\text{ini}}\rangle $ that corresponds to one of the following basis vectors:
 
$\bullet $ Domain wall, $|\rm{DW}\rangle = | \uparrow \uparrow \uparrow \ldots \downarrow \downarrow \downarrow \rangle$, 

$\bullet $ N\'eel state, $ |\rm{NS}\rangle= | \uparrow \downarrow \uparrow \downarrow \ldots  \uparrow \downarrow \uparrow \downarrow \rangle$, 

$\bullet $ Pairs of parallel spins, $ |\rm{PP}\rangle=| \downarrow \uparrow  \uparrow  \downarrow \downarrow \uparrow  \uparrow \downarrow \downarrow \ldots   \rangle$. 

These states are, in principle, accessible to experiments in optical lattices~\cite{Simon2011}. The preparation of a sharp domain wall requires a magnetic field gradient as realized in~\cite{Weld2009}, and the possibility for achieving the N\'eel state has been discussed in~\cite{Koetsier2008,Mathy2012}. In addition to the experimental motivation, these states are chosen for their enhanced effects of the Ising interaction and NNN couplings. They all belong to the same subspace ${\cal S}^z =0$ with dimension ${\cal D}=\begin{pmatrix}L \\ L/2\end{pmatrix}$.

We also analyze unpolarized random initial states:

$\bullet $ in the subspace ${\cal S}^z = 0$,  $|\xi_{{\cal S}^z = 0} \rangle$,

$\bullet $ in the whole Hilbert space, $|\xi_{2^L} \rangle$. 

They are random superpositions of the site-basis vectors. The probability amplitude for each of these site-basis vectors has the same modulus $1/\sqrt{{\cal D}}$ and a random phase $e^{i 2\pi\varphi}$, where $ \varphi $ is a uniformly distributed random variable in $ [0,1) $. These random states manifest thermal features since the evaluation of local observables yields the same outcomes as that for a mixed state of infinite temperature. In particular, they exhibit a self-averaging property that can be employed to evaluate ensemble spin dynamics~\cite{Alvarez2008}, and they have already been used to compute high-temperature correlation functions~\cite{Alvarez2010,Zangara2012}. Since they are already at thermal equilibrium, they can be used to set the minimum amplitude of the time fluctuations.

\subsection{Few-body observables}

We study the relaxation and time fluctuations of the following few-body observables.

$\bullet$ Kinetic energy,
\begin{eqnarray}
\widehat{{\rm KE}}&=& \sum_{n=1}^{L-1} J \left(\hat{S}_n^x \hat{S}_{n+1}^x + \hat{S}_n^y \hat{S}_{n+1}^y  \right) 
\nonumber \\
&+& \lambda \sum_{n=1}^{L-2} J \left(\hat{S}_n^x \hat{S}_{n+2}^x + \hat{S}_n^y \hat{S}_{n+2}^y  \right).
\end{eqnarray}

$\bullet$ Interaction energy,
\begin{equation}
\widehat{{\rm IE}}= \sum_{n=1}^{L-1} J \Delta \hat{S}_n^z \hat{S}_{n+1}^z +
\lambda \sum_{n=1}^{L-2} J \Delta \hat{S}_n^z \hat{S}_{n+2}^z.
\end{equation}
The time fluctuations for $\widehat{{\rm KE}}$ and $\widehat{{\rm IE}}$ are the same, since the two observables add up to the constant total energy, so we show results only for $\widehat{{\rm KE}}$. 

$\bullet $ Spin-spin correlations in the $z$ and $x$ direction,
\begin{equation}
\hat{C}^{z(x)}_{nm}= \hat{S}_n^{z(x)} \hat{S}_{m}^{z(x)}.
\end{equation}
We present results for $ n=L/2 $ and $ m=L/2+1$, but studied also $ m=L/2+2 $ and $ L/2+3 $. Since the interactions considered here are short-range, these correlations decay with the distance between spins $n$ and $ m $. The restriction to sites in the middle of the chain is to minimize boundary effects. 

$\bullet $ Structure factors in  $z$ and $x$,
\begin{equation}
\hat{s}_f^{z(x)}(k)=\frac{1}{L} \sum_{n,m=1}^L e^{i k (n-m) } \hat{S}_n^{z(x)} \hat{S}_{m}^{z(x)}.
\end{equation}
They are the Fourier transform of the spin-spin correlations with $k=2\pi p/L$ and $p$ as an integer, $p=1,\ldots,L$.  For the fluctuations, we present results only for $k=\pi$ since this momentum exists for all system sizes considered here, $10\leq L \leq 22$. We have also studied the sum over all $k's$ and the results are qualitatively very similar. The time evolution, however, shows visible differences associated with the value of $k$. This is discussed in Sec.~\ref{sec:relaxation}.

\subsection{ Numerical method} 

Exact diagonalization is employed for describing static properties of the system with $L=16$ (${\cal S}^z = 0$) and $L=18$ (${\cal S}^z = -3$). The dynamics, however, involves chains with up  to $L=24$, which rules out the possibility of using full exact diagonalization. Instead, the time evolution of the pure states defined above is evaluated by means of a fourth order Trotter-Suzuki (TS) decomposition~\cite{DeRaedt1999,DeRaedt2000}. 

The TS method is a standard strategy in
which an evolution operator $\hat{U}(\delta t)=\exp[- i \hat{H}  \delta t ]$ is approximated by an
appropriate sequence of partial evolution operators in the form $\tilde{U}(\delta t)=
\prod_{k} \exp[-i \hat{H}_{k} \delta t ]$. Here, $ \{ \hat{H}_{k} \} $ is a set of Hermitian operators
properly chosen for the purpose of providing a simple and efficient implementation of
each partial evolution. We choose $\hat{H}_k$ so that it only contains a two-spin operation in a given direction ($x$, $y$  or $z$), {\em e.g.} $\hat{S}_{n}^{y} \hat{S}_{m}^{y}$ \cite{DeRaedt1999,DeRaedt2000}. This avoids manipulating and diagonalizing
the total Hamiltonian $\hat{H}$.
The evaluation of the dynamics for an arbitrary finite time $t$ requires the
successive application of the steplike evolutions $\tilde{U}(\delta t)$. Even though the
approximated dynamics always remains unitary, the accuracy of the approximation
relies on the TS time step $\delta t $ being sufficiently small. In particular, it must be
much smaller than the maximum local time scale, which is determined by the local
second moment of $\hat{H}$. We have tuned the TS time step so that, for the largest system sizes
$L = 22, 24$, relative errors bounds are estimated at $10^{-6}$ for maximum evolution
times of $Jt = 5 \times 10^3$.

We implemented the TS method on general purpose graphical processing units. Such hardware enables a massive
parallelization scheme in the site basis, which yields a substantial speedup of our
simulations~\cite{Dente2013}. 
We stress that our approach is exact within the TS approximation and it does not require any specific
symmetry to be assumed. This means
that there are no truncations of the Hilbert space. 
Since the full Hilbert space is
available, there are no truncation errors that would drastically limit the access of long-time
asymptotics. This is rather crucial, as it is often the major obstacle when
addressing long-time dynamics in interacting many body systems using standard
strategies, such as time-dependent density-matrix renormalization-group \cite{Prosen_PRE2007} and tensor network techniques \cite{Banuls2012}. 

\section{Spectrum and Energy Shell}

The initial state $|\text{ini}\rangle$ evolves according to $|\Psi(t)\rangle=\sum_{\alpha} C_{\alpha}^{\text{ini}} e^{-i  E_{\alpha} t} |\alpha \rangle$, where  $C_{\alpha}^{\text{ini}} = \langle \alpha| \text{ini} \rangle$ and $E_{\alpha}$ and $|\alpha \rangle$ are the eigenvalues and eigenstates of the Hamiltonian dictating the dynamics of the system. 

The expectation value of an observable $O$ at time $t$ is given by
\begin{eqnarray}
&&\langle \hat{O}(t) \rangle = \langle \Psi(t) |\hat{O} |\Psi(t) \rangle 
\nonumber \\
&&= \sum_{\alpha} |C_{\alpha}^{\text{ini}}|^2   O_{\alpha \alpha} + \sum_{\alpha \neq \beta} C_{\alpha}^{\text{*ini}}  C_{\beta}^{\text{ini}}   O_{\alpha \beta} e^{i (E_{\alpha} - E_{\beta}) t} ,
\label{Eq:Obs}
\end{eqnarray}
where $O_{\alpha \beta} =\langle \alpha| \hat{O} |\beta \rangle $ are the matrix elements of the observable.
The variance of the temporal fluctuations of the observable about its equilibrium value corresponds to
\begin{eqnarray}
&&\sigma^2_O = \overline{ |\langle O(t) \rangle - \overline{\langle O(t) \rangle}|^2}  
\label{Eq:sigma}  \\
&&= 
\mathop{\sum_{\alpha \neq \beta} }_{\gamma \neq \delta} 
C_{\alpha}^{\text{*ini}}  C_{\beta}^{\text{ini}} C_{\gamma}^{\text{*ini}}  C_{\delta}^{\text{ini}}
 O_{\alpha \beta} O^{\dagger}_{\gamma \delta} 
\overline{e^{i (E_{\alpha} - E_{\beta}+E_{\gamma} - E_{\delta}) t}} 
\nonumber
\end{eqnarray}
where $\overline{O} = T^{-1} \int_{0}^T O(t) dt$ is the time average over the interval $[0,T]$.

Under the condition of nondegenerate energy gaps,
\begin{eqnarray}
&& E_{\alpha} = E_{\beta} \hspace{0.2 cm } \text{and}  \hspace{0.2 cm } E_{\delta} = E_{\gamma} \nonumber \\
E_{\alpha} - E_{\beta} = E_{\delta} - E_{\gamma}  \Rightarrow && \text{or} \nonumber \\
&& E_{\alpha} = E_{\delta} \hspace{0.2 cm } \text{and}  \hspace{0.2 cm } E_{\beta} = E_{\gamma} 
\label{gap}
\end{eqnarray}
and for $T\rightarrow \infty$, it has been shown that~\cite{Reimann2008,Short2011} 
\begin{equation}
\sigma^2_O \leq (O_{\text{max}} - O_{\text{min}})^2 \text{Tr}[\overline{\rho}^2 ] =\frac{(O_{\text{max}} - O_{\text{min}})^2 }{\text{IPR}^{\text{ini}} } ,
\label{bound}
\end{equation}
where $O_{\text{max(min)}}$ is the maximum (minimum) eigenvalue of the operator $\hat{O}$, $\overline{\rho}=\sum_{\alpha} |C_{\alpha}^{\text{ini}} |^2 |\alpha \rangle \langle \alpha |$ is the diagonal density matrix~\cite{pure}, and
\begin{equation}
\text{IPR}^{\text{ini}} =\frac{1}{\sum_{\alpha} | C_{\alpha}^{\text{ini}}|^4}
\end{equation}
is the inverse participation ratio of the initial state in the energy eigenbasis. The bound above has been further improved and the condition of nondegenerate gaps was substituted by not too many~\cite{Short2012}. 

IPR measures the level of delocalization of a state in a certain basis. In full random matrices, the eigenstates are maximally delocalized. For random matrices from a Gaussian Orthogonal Ensemble~\cite{Guhr1998}, it is found that $\text{IPR} \sim {\cal D}/3$ \cite{Izrailev1990}, while for a Gaussian Unitary Ensemble~\cite{Guhr1998}, $\text{IPR} \sim {\cal D}/2$ .  Here,  none of the initial states taken from basis vectors reach such large values for $\text{IPR}^{\text{ini}}$, which is not surprising, since they are not eigenstates from random matrices. In contrast, the thermal states, indeed, have $\text{IPR}^{\text{ini}}\sim {\cal D}/2$. For few-body observables, a delocalized initial state with $\text{IPR}^{\text{ini}} \propto {\cal D}$ leads to the exponential decay of $\sigma _O$ with system size, since ${\cal D}$ grows exponentially with $L$. 

Below, we first present results for the level spacing distribution and number of degenerate energy gaps. They reinforce the expectation of an exponential decay with $L$ for the time fluctuations of few-body observables in chaotic spin-1/2 systems and integrable ones with $0<\Delta \leq 1$. Next, we present the level of delocalization of the initial states for these systems. They set the bounds for $\sigma_O$ in Eq.~(\ref{bound}) and help justify the value of the coefficient of the exponential decay found numerically in the following section. We notice that, even though the bound also depends on the range of the eigenvalues of $O$, the observables considered here hardly affect the value of the coefficient (see Sec.~IV).

\subsection{Spectrum}
\label{sec:spectrum}

\begin{figure}[htb]
\centering
\includegraphics[width=0.45\textwidth]{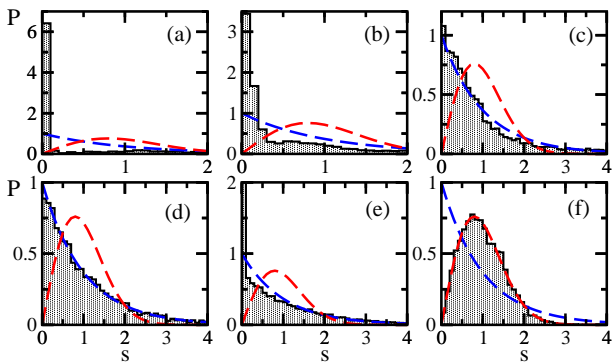}
\caption{(Color online) Level spacing distribution for a single subspace: $L=18$, ${\cal S}^z=-3$, and eigenstates with even parity. For comparison, the Poisson and Wigner-Dyson distribution are shown with dashed lines. (a) -- (e)  have $\lambda= 0$  and $\Delta=0.0, 10^{-3}, 10^{-2}, 0.1$, and 0.5, respectively. (f) $\Delta=0.5$ and $\lambda=1$.}
\label{fig:WD}	
\end{figure}

Absence of degeneracies goes hand in hand with chaotic systems, where the energy levels are correlated and crossings are avoided. The distribution $P(s)$ of spacings $s$ between neighboring unfolded energy levels is Wigner-Dyson (WD) \cite{HaakeBook,Guhr1998,ReichlBook}.  The exact shape of this distribution depends on the symmetries of the system. In the case of time-reversal symmetry, we have $P_{WD}(s)=(\pi s/2) \exp(-\pi s^2/4)$. In integrable systems, the eigenvalues tend to cluster and are not prohibited from crossing. The level spacing distribution is Poissonian, $P_P(s) = \exp(-s)$. As $\lambda$ increases from zero in Eq.~(\ref{ham}), a WD-distribution is eventually obtained, as shown in the panel (f) of Fig.~\ref{fig:WD} \cite{crossover}.  We show results for the subspace ${\cal S}^z=-3$ and $\Delta \neq 1$, to avoid symmetries associated with global $\pi$-rotation around $x$ and ${\cal \hat{S}}^2$. Only parity needs to be taken into account, so the statistics is still very good; for $L=18$ and even parity we have $\sim 10^4$ energy levels.

In the presence of too many degeneracies or in localized systems, one observes deviations from  $P_P(s)$  with the emergence of the Shnirelman peak~\cite{Shnirelman1975,Chirikov1995,Frahm1997}. This is seen in panel (a) of  Fig.~\ref{fig:WD} where we show the level spacing distribution for the XX-model ($\Delta = \lambda= 0$). The number of small spacings there goes much beyond the Poisson distribution. However, as $\Delta$ increases (the XXZ model), the excessive degeneracies rapidly fade away [compare the distribution for $\Delta=10^{-3}$ in panel (b) with that for $\Delta=10^{-2}$ in panel (c)]. 

Panels (d) and (e) show results for the NN system with $\Delta=0.1$ and 0.5, respectively. Notice that, for the special value 1/2, the form of the distribution also departs from $P_P(s)$, although the Poisson distribution is recovered by changing it slightly, for example, by using $\Delta=0.48$. 

In addition to no (few) energy degeneracies, $\text{gap}_{\alpha\beta}=|E_{\alpha} - E_{\beta}| \neq 0$, a main condition for the exponential decay of the temporal fluctuations in Eq.(\ref{bound}) is the existence of no (few) degenerate energy gaps, $\delta \text{gap} = |\text{gap}_{\alpha'\beta'} - \text{gap}_{\alpha\beta}| \neq 0$. In Table \ref{table:gap} , we compare the total number of energy differences where $\text{gap}_{\alpha\beta}<10^{-8}$ and total number of gap differences where $\delta \text{gap} <10^{-8}$ for systems with NN couplings and $\Delta=0, 0.01, 0.1$ and 0.5, as well as for the chaotic system with $\lambda=1 $ and $\Delta=0.5$.
\begin{table}[h]
\caption{Total number of energy differences where $\text{gap}_{\alpha\beta}<10^{-8}$ and of gap differences where $\delta \text{gap} <10^{-8}$; $L=15$; ${\cal S}^z=-3$, eigenstates with even parity.}
\begin{center}
\begin{tabular}{|l|c|c|}
\hline 
  &    $\text{gap}_{\alpha\beta}<10^{-8}$  & $\delta \text{gap} <10^{-8}$\\
  \hline
$\hat{H}_{\Delta=0,\lambda=0}$  & $2 088$ &   $336\, 508\, 464$  \\
\hline
$\hat{H}_{\Delta=0.01,\lambda=0}$   & $0$ &  $4\, 202$ \\
\hline
$\hat{H}_{\Delta=0.1,\lambda=0}$   & $0$ & $4\, 020$    \\
\hline
$\hat{H}_{\Delta=0.5,\lambda=0}$   & $192$ & $347\, 844$    \\
\hline
$\hat{H}_{\Delta=0.5,\lambda=1}$  & $0$   &  $2\, 632$   \\
\hline
\end{tabular}
\end{center}
\label{table:gap}
\end{table}

As seen in Table \ref{table:gap}, the number of energy and gap degeneracies in the XX-model (first row) is enormous. It drops abruptly with the introduction of the Ising interaction, even for strengths as low as $\Delta=0.01$. For $L=15$, $\delta \text{gap}$ is 5 orders of magnitude smaller for the integrable models with anisotropy (second and third rows) than for $\Delta=0$.  For these cases, the number of gap degeneracies  is comparable to that in the chaotic model (last row). This justifies the expectation for an exponential decay of the time fluctuations with $L$ for integrable systems with $0<\Delta \leq 1$. 

Notice, however, the special behavior of the XXZ model with $\Delta=1/2$ (fourth row). This point shows energy degeneracies, as also seen in Fig.~\ref{fig:WD} (e), and a large number of gap degeneracies, even though $\delta \text{gap}$  is still 3 orders of magnitude smaller than for the XX model.
Our choice of $\Delta=1/2$ in the numerical studies of Sec.~\ref{sec:fluctuations} is, therefore, not arbitrary. If an exponential behavior is observed even for this particular case, then it is certain to occur for the other gapless XXZ chains.

\subsection{Energy shell}
\label{sec:shell}

Since our systems only have two-body interactions, a maximum delocalized $|\text{ini}\rangle $ is the one that fills the energy shell ergodically. The energy shell is a Gaussian centered at the energy of the initial state, 
\begin{equation}
E_{\text{ini}} =\sum_{\alpha} |C_{\alpha}^{\text{ini}}|^2 E_{\alpha} = H_{\text{ini},\text{ini}}
\label{Eini}
\end{equation}
with squared width,
\begin{equation}
\delta E_{\text{ini}}^2 = \sum_{\alpha} |C_{\alpha}^{\text{ini}} |^2 (E_{\alpha} - E_{\text{ini}})^2 =\sum_{j\neq \text{ini}} |H_{\text{ini},j}|^2.
\label{deltaE}
\end{equation}
The last equality in the two equations above holds when the initial state is one of the basis vectors. In this case, we do not need to diagonalize the Hamiltonian to obtain the energy shell, we simply need the elements $H_{i,j}$ of the Hamiltonian matrix~\cite{Flambaum1997,Santos2012PRE}. 
The diagonal elements, which determine $E_{\text{ini}}$, only depend on the NN and NNN Ising terms. Pairs of parallel NN and NNN spins in the $z$ direction contribute positively to the energy of the state, whereas pairs of antiparallel spins contribute negatively.

We can see, for instance, that the domain wall state has 
\begin{equation}
E_{\text{ini}}^\text{DW} =\frac{J\Delta}{4} [(L-3) + (L-6)\lambda],
\label{Eq:DW}
\end{equation}
where both NN and NNN Ising interactions contribute with positive signs to the energy, and
\begin{equation}
\delta E_{\text{ini}}^\text{DW} = \frac{J}{2}\sqrt{1+2\lambda^2}.
\label{Eq:DWwidth}
\end{equation}
Notice that the width of the shell for this state does not depend on the system size.

\begin{figure}[htb]
\centering
\includegraphics*[width=0.45\textwidth]{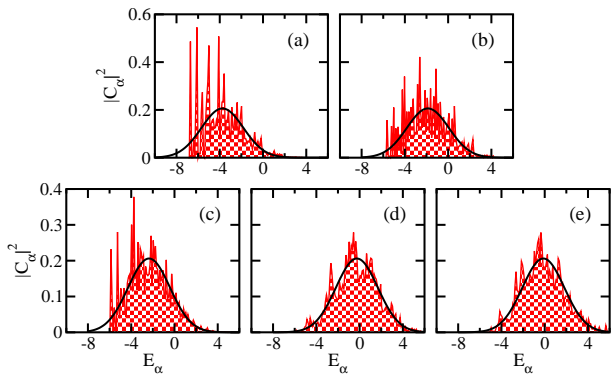}
\caption{(Color online) Distribution of the weights of the initial N\'eel state in the energy representation, $L=16$, ${\cal S}^z=0$. The Hamiltonians and $E_{\text{ini}}$ are as follows: (a) $\hat{H}_{\Delta=1,\lambda=0}$ and -3.750; (b) $\hat{H}_{\Delta=0.5,\lambda=0}$ and -1.875; (c) $\hat{H}_{\Delta=1,\lambda=0.4}$ and -2.350; (d) $\hat{H}_{\Delta=1,\lambda=1}$ and -0.250; and (e)  $\hat{H}_{\Delta=0.5,\lambda=1}$ and -0.125. The solid line corresponds to the energy shell: Gaussian of width $\delta E_{\text{ini}}=1.936$.}
\label{fig:shell}	
\end{figure}

Figure~\ref{fig:shell} shows the distribution of $|C_{\alpha}^{\text{ini}} |^2$ in the eigenvalues $E_{\alpha}$ for the N\'eel state for the Hamiltonians with $\lambda=0$ ($\Delta=1,0.5$), $\lambda =0.4 (\Delta=1$), and $\lambda=1$ ($\Delta=1,0.5$). The width of the energy shell is the same for all cases, because the direct coupling between $|\text{NS}\rangle$ and the other site-basis vectors is only due to the NN flip-flop term, so 
\begin{equation}
\delta E_{\text{ini}}^\text{NS}=\frac{J}{2} \sqrt{L-1}.
\end{equation}
The number of contributing levels, on the other hand, differs significantly from one model to the other. As $\Delta$ decreases and $\lambda$ increases, 
\begin{equation}
E_{\text{ini}} ^\text{NS}=  \frac{J\Delta}{4}  [ -(L-1) + (L-2)\lambda ]
\label{NeelE}
\end{equation}
approaches the middle of the spectrum, where the density of states is large, and so the energy shell gets better filled, as in Fig.~\ref{fig:shell} (e). Closer to the edges of the spectrum, the distribution is less homogenous, spiky, and asymmetric, as in Fig.~\ref{fig:shell} (a).

\begin{figure}[htb]
\centering
\includegraphics*[width=0.45\textwidth]{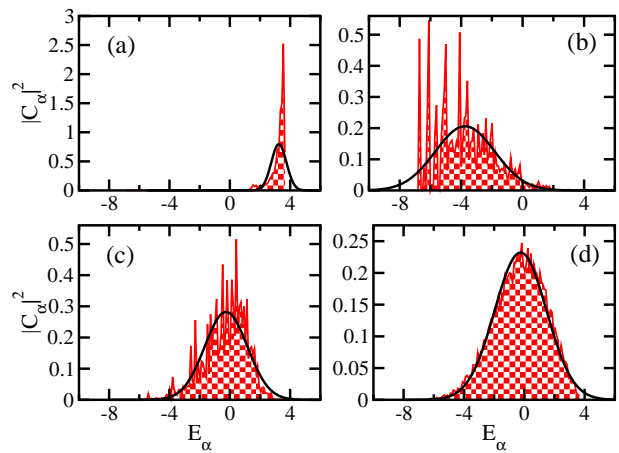}
\caption{(Color online) Top panels: Distribution of the weights of the initial state in the energy representation for $\hat{H}_{\Delta=1,\lambda=0}$, $L=16$, and ${\cal S}^z=0$. (a) $|\text{DW}\rangle$,  $E _{\text{ini}}=3.250$ and $\delta E_{\text{ini}}=0.500$; (b)  $|\text{NS}\rangle$, $E _{\text{ini}}=-3.750$ and $\delta E_{\text{ini}}=1.936$; (c) $|\text{PP}\rangle$, $E _{\text{ini}}=-0.250$ and $\delta E_{\text{ini}}=1.414$; and (d) $|\xi_{{\cal S}^z = 0} \rangle$, $E _{\text{ini}}=-0.246 $ and $\delta E_{\text{ini}}= 1.719 $.}
\label{fig:shellH1}	
\end{figure}

The level of delocalization of the initial state depends on the combined relationship between $|\text{ini}\rangle $ and $\hat{H}$. A better notion of the role of the initial state may be gained from Fig.~\ref{fig:shellH1}, where we fix the Hamiltonian and change $|\text{ini}\rangle $. We select the most restrictive case among the $\hat{H}$'s with $0<\Delta\leq1$, that is the integrable isotropic Hamiltonian $\hat{H}_{\Delta=1,\lambda=0}$. The distribution of the components of the initial state goes as follows. The domain wall and the N\'eel state are both at the edges of the spectrum, the first at very high energy and the second at very low energy. The distribution for $|\text{DW}\rangle$ is narrow and spiky [Fig.~\ref{fig:shellH1} (a)] and $\delta E_{\text{ini}}$ is much smaller than $\omega$ (cf. Table~\ref{table:DOS} and caption of Fig.~\ref{fig:shellH1}). The distribution for $|\text{NS}\rangle$ is broad, in fact $\delta E_{\text{ini}} \gtrsim \omega$, but the shell is not well filled. This is noticed from the many spikes in Fig.~\ref{fig:shellH1} (b) and also from the low value of the $\text{IPR}^{\text{ini}}$ in Table~\ref{table:IPR}. The distribution for $|\text{PP}\rangle$, which is a state close to the middle of the spectrum, is relatively broad, $\delta E_{\text{ini}} \lesssim \omega$, and the shell is relatively well filled [Fig.~\ref{fig:shellH1} (c)]. It is only when the initial state is one of the thermal states, $|\xi_{{\cal S}^z = 0} \rangle$ or $|\xi_{2^L} \rangle$, that the distribution becomes independent of the Hamiltonian, the energy shell being filled ergodically for any $\hat{H}$ and $\delta E_{\text{ini}} \sim \omega$. This is illustrated in Fig.~\ref{fig:shellH1} (d) for $|\xi_{{\cal S}^z = 0} \rangle$. 

\begin{table}[h]
\caption{Inverse participation ratio of the initial states corresponding to site-basis vectors for $L=12,14,16$ in ${\cal S}^z=0$.}
\begin{center}
\begin{tabular}{|c|c|c|c|}
\hline 
  &  IPR$^{\text{ini}}_{L=12}$ & IPR$^{\text{ini}}_{L=14}$ & IPR$^{\text{ini}}_{L=16}$  \\
 \hline
 \hline
 $\hat{H}_{\Delta=1.5,\lambda=0}$ &    &    &   \\
 \hline
 $|\text{DW}\rangle$ &  2.862  &  1.436 &  1.432 \\
$|\text{NS}\rangle$ & 15.782    &  23.865 & 35.981  \\
$|\text{PP}\rangle$ &  22.870  &  39.528  & 64.051  \\
  \hline
  \hline
$\hat{H}_{\Delta=1,\lambda=0}$ &    &    &  \\
\hline
$|\text{DW}\rangle$ &  16.986  & 24.541   & 34.858  \\
$|\text{NS}\rangle$ & 24.580  & 42.003   &  72.153    \\
$|\text{PP}\rangle$ & 45.814   & 95.851   &  200.570   \\
\hline
\hline
$\hat{H}_{\Delta=0.5,\lambda=0}$ &    &   &       \\
\hline
$|\text{DW}\rangle$ & 37.259   & 63.718   &  104.334     \\
$|\text{NS}\rangle$ & 38.575   &  70.555  &  129.782     \\
$|\text{PP}\rangle$ &  50.697  & 109.737   &  241.171   \\
\hline
\hline
$\hat{H}_{\Delta=1,\lambda=0.4}$ &    &     &    \\
\hline
$|\text{DW}\rangle$ & 15.643   & 22.593   & 31.948     \\
$|\text{NS}\rangle$ & 64.316   & 147.957   &  336.776    \\
$|\text{PP}\rangle$ & 73.936   & 218.272   &  592.725    \\
\hline
\hline
$\hat{H}_{\Delta=1,\lambda=1}$ &    &     &    \\
\hline
$|\text{DW}\rangle$ &  14.380  & 20.521   & 28.690  \\
$|\text{NS}\rangle$ &  168.345  & 514.499   &  1805.249   \\
$|\text{PP}\rangle$ &  31.851  &  68.373  &   129.883   \\
\hline
\hline
$\hat{H}_{\Delta=0.5,\lambda=1}$ &    &     &    \\
\hline
$|\text{DW}\rangle$ & 50.567   & 123.785   &  368.140  \\
$|\text{NS}\rangle$ &  158.029  &  548.877  &  2071.923   \\
$|\text{PP}\rangle$ &  77.661  & 228.241   &  586.557   \\
\hline
\end{tabular}
\end{center}
\label{table:IPR}
\end{table}

The two factors together, broadening and filling of the energy shell, improve from (a) to (d) in Fig.~\ref{fig:shellH1} and are reflected in the values of $\text{IPR}^{\text{ini}}$ in Table~\ref{table:IPR}. The domain wall is the most localized of the states. For $\hat{H}_{\Delta=1,\lambda=0}$, $\text{IPR}^{\text{ini}}$ then increases from $|\text{NS}\rangle$ to $|\text{PP}\rangle$, but of course never reaches the level of delocalization of eigenstates from random matrices. Only for the thermal states, $\text{IPR}^{\text{ini}} \sim {\cal D}/2$ for any of the Hamiltonians considered (not shown). In this same order, we expect the decay of the fluctuations with $L$ and the time evolution of the observables to become faster. 

Notice from Table~\ref{table:IPR} that the level of delocalization of $|\text{PP}\rangle$ is larger than that of $|\text{NS}\rangle$ when $\lambda$ is small, but this changes for $\lambda=1$. The NNN Ising term contributes negatively to $|\text{PP}\rangle$, so it pushes the state away from the middle of the spectrum towards low energies, this being accentuated for large $\lambda$. As a result, at the isotropic point, $\text{IPR}^{\text{ini}}$ for $|\text{PP}\rangle$ is larger for weak chaos ($\lambda =0.4$) than for strong chaos ($\lambda=1$). This is surprising, because the majority of the states get more delocalized as the level of chaoticity increases. In contrast, for the N\'eel state, the NNN Ising interaction adds energy and counterbalances the effects of the NN term, which is negative  [Eq.(\ref{NeelE})], so larger $\lambda$ implies a state closer to the middle of the spectrum and therefore more spread.

We are not able to perform a scaling analysis with the values of $\text{IPR}^{\text{ini}}$, because only three system sizes are available. We then look for indications of the exponential decay of the time fluctuations with $L$ directly in the numerical studies of the observables. However, some observations can already be made at this point. From the definition of the thermal states, it is clear that $\text{IPR}^{\text{ini}}$ grows exponentially with $L$ and so will the reciprocal of $\sigma_O$ for few-body observables [Eq.(\ref{bound})]. As seen in Table~\ref{table:IPR}, the value of the ratio $\text{IPR}^{\text{ini}}/{\cal D}$ for the N\'eel state in the strongly chaotic Hamiltonian $\hat{H}_{\Delta=0.5,\lambda=1}$ is also constant ($\sim1/6$), so here again the exponential behavior of  $\sigma_O$ is guaranteed. For the other initial states and $\hat{H}$'s, $\text{IPR}^{\text{ini}}$ grows slower than ${\cal D}$, especially for $|\text{DW}\rangle $ in the isotropic points, but it may as well be an exponential growth. The only clear exception is the domain wall in the gapped phase ($\Delta=1.5$), which as expected, further localizes as $L$ increases. A discussion about the relaxation process of this state in the gapped and gapless phases is presented in Sec.~\ref{sec:relaxation}.

\section{Numerical Results for the Time Fluctuations} 
\label{sec:fluctuations}

Our numerical results indeed suggest that the standard deviation of the time fluctuations for chaotic and integrable systems with $0<\Delta \leq 1$ decay exponentially with system size $\sigma_O \propto e^{ - \kappa L}$. The value of the coefficient $\kappa$ of this decay increases significantly with the level of delocalization of the initial state and, when comparing observables, it is usually slightly larger for $\widehat{{\rm KE}}$. In order to elucidate the fluctuations decay law, we analyze each observable, initial state, and Hamiltonian in log-linear and log-log scales. Linear fittings in these scales enable a quantitative comparison between the two possibilities, based on the standard coefficient of determination $R^2$. 

\begin{figure}[htb]
\centering
\includegraphics[width=0.45\textwidth]{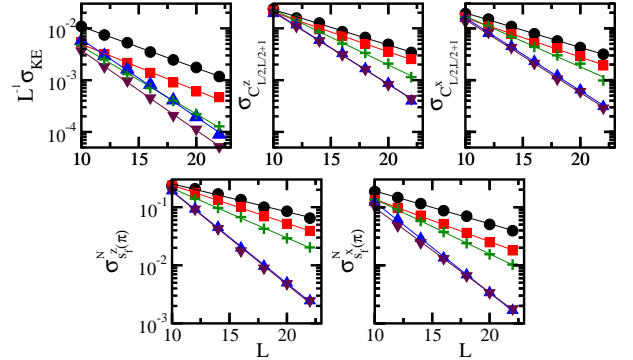}
\caption{(Color online) Logarithmic plot of the standard deviation of the time fluctuations of different observables vs $L$ for (circles) $\hat{H}_{\Delta=1,\lambda=0}$,   (squares) $\hat{H}_{\Delta=0.5,\lambda=0}$,  (plus) $\hat{H}_{\Delta=1,\lambda=0.4}$,  (up triangle) $\hat{H}_{\Delta=1,\lambda=1}$, and  (down triangle) $\hat{H}_{\Delta=0.5,\lambda=1}$. $|\text{ini} \rangle$ is the N\'eel state; $[100,500]$ is the time interval for the averages; and $\sigma_O^N=\sigma_O/\overline{O}$. The solid lines correspond to logarithmic fits.}
\label{fig:SigPsi2}	
\end{figure}

In Fig.~\ref{fig:SigPsi2}, results for $\sigma_O$  are shown for different observables in the case where $|\text{ini}\rangle$ is the N\'eel state. The dispersion is computed in a time interval after the observables reached a stationary state. The exponential decay with $L$ is evidenced by excellent linear fits for the log-linear plots of integrable and chaotic Hamiltonians. When comparing with power-law fittings, the values of $R^2$ are systematically worse, although not overwhelmingly worse [some examples are given in Sec.~\ref{sec:compare}]. There is just one case, for $\widehat{{\rm KE}}$ and $\hat{H}_{\Delta=0.5,\lambda=0}$, where $R^2$ for the power law barely exceeds the exponential fitting. The irrefutable rejection of a power-law behavior would require system sizes beyond the ones considered here, $10\leq L \leq 22$. Nevertheless, we emphasize that the exponent $b$ in the power-law fitting, $\sigma_O \propto L^{-b}$, is always much larger than the value 0.5 found in systems of quasifree particles~\cite{Venuti2013,Cassidy2011,Gramsch2012,HeARXIVdisorder}. For the N\'eel state and the observables studied in Fig.~\ref{fig:SigPsi2}, the smallest factor was $b\sim2$, which was found for $\hat{H}_{\Delta=1,\lambda=0}$ and $\hat{s}_f^{z(x)}(\pi)$.

The coefficient $\kappa$ in the exponential fittings of Fig.~\ref{fig:SigPsi2} decreases with $\Delta$ and increases with $\lambda$.  As the anisotropy increases,  $| \text{ini} \rangle$ becomes less spread in the energy representation, as seen in Fig.~\ref{fig:shell} and Table~\ref{table:IPR}. The excitations in the system lose mobility as it  passes the isotropic point ($\Delta=1$), where total spin is conserved, and then enters the gapped phase   ($\Delta >1$), where well separated bands of energies are formed \cite{JoelARXIV}. On the other hand, as $\lambda$ increases from zero, the crossover to chaos takes place, thus favoring the delocalization of the N\'eel state. The competition between NN and NNN interactions brings this state close to the middle of the spectrum [Eq.~(\ref{NeelE})]. The value of $\kappa$ reflects the width of the energy shell as well as its filling. For the particular case of $|\text{NS}\rangle$, where the width of the shell is always the same (cf. Fig.~\ref{fig:shell}), it is the filling of the shell that leads to the different coefficients in Fig.~\ref{fig:SigPsi2}. In the figure, the smallest value is $\kappa \sim 0.11$ for  $\hat{H}_{\Delta=1,\lambda=0}$ and $\hat{s}_f^z(\pi)$ and the largest is $\kappa \sim 0.37$ for $\hat{H}_{\Delta=0.5,\lambda=1}$ and $\hat{s}_f^z(\pi)$, as seen in Table~\ref{table:kappaPsi2}.

\begin{table}[h]
\caption{Coefficient $\kappa$ in the exponential fittings $\sigma_O \propto e^{ - \kappa L}$ of Fig.~\ref{fig:SigPsi2}  for $|\text{ini}\rangle = |\text{NS}\rangle$.}
\begin{center}
\begin{tabular}{|l|c|c|c|}
  \hline
  \multicolumn{4}{|c|}{$\kappa$ for $|\text{NS}\rangle$} \\
\hline 
  $\hat{O}=$&  $\widehat{KE}$ &  $\hat{C}_{L/2,L/2+1}^z$ & $\hat{s}_f^z(\pi)$ \\
  \hline
$\hat{H}_{\Delta=1,\lambda=0}$ & 0.184 &  0.157  &  0.111     \\
\hline
$\hat{H}_{\Delta=0.5,\lambda=0}$ & 0.206 &  0.175  & 0.151      \\
\hline
$\hat{H}_{\Delta=1,\lambda=0.4}$ & 0.301 & 0.246 &  0.196      \\
\hline
$\hat{H}_{\Delta=1,\lambda=1}$ & 0.345 & 0.324  &  0.366      \\
\hline
$\hat{H}_{\Delta=0.5,\lambda=1}$ & 0.354 & 0.320   &   0.369     \\
\hline
\end{tabular}
\end{center}
\label{table:kappaPsi2}
\end{table}

The N\'eel state behaves as a thermal state for the chaotic Hamiltonian $\hat{H}_{\Delta=0.5,\lambda=1}$. It fills the energy shell very well and $\text{IPR}^{\text{ini}} \sim {\cal D}/6$ (cf. Fig.~\ref{fig:shell} and Table~\ref{table:IPR}). This explains the value $\kappa \sim 0.35$, which is the same as that obtained for the initial random states, $|\xi_{{\cal S}^z = 0} \rangle$ and $|\xi_{2^L} \rangle$. As mentioned earlier, these latter states fill the energy shell ergodically for any of the Hamiltonians, therefore, their time fluctuations are the minimum possible ones. According to Eq.~(\ref{bound}) and using ${\cal D}=2^L$ from $|\xi_{2^L} \rangle$, we see that  $\sigma^2\sim 2^{-L}$, which yields the value of $\kappa=\frac{1}{2}\log 2\approx 0.35$.  Therefore, at least when the initial state fills the energy shell, the agreement between the analytical prediction and our numerical results is excellent.

Contrary to $|\text{NS}\rangle$, the domain wall does not reach high levels of delocalization, since it is far from the middle of the spectrum. The values of $\kappa$ are significantly smaller, especially at the critical point $\Delta=1$. Even for $\hat{H}_{\Delta=0.5,\lambda=1}$, $\kappa$ does not reach the maximum 0.35. It gets close to it for $\widehat{{\rm KE}}$ ($\kappa \sim 0.33$), but it does not pass 0.28 for the other observables. 

\begin{figure}[htb]
\centering
\includegraphics[width=0.45\textwidth]{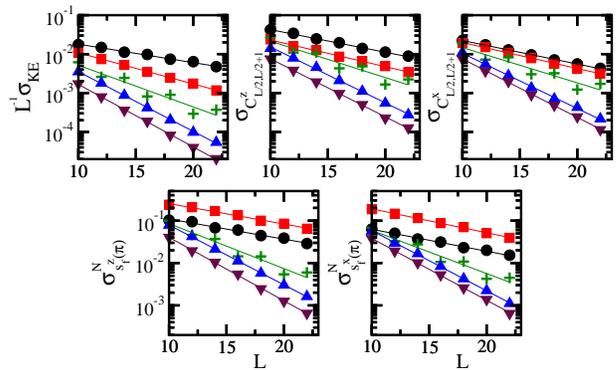}
\caption{(Color online)  Logarithmic plots of the standard deviation of the time fluctuations for different observables vs $L$ for  (circles) $|\text{DW}\rangle$,  (squares) $|\text{NS}\rangle$,  (plus) $|\text{PP}\rangle$,  (up triangle) $|\xi_{{\cal S}^z = 0} \rangle$, and  (down triangle) $|\xi_{2^L} \rangle$.  The solid lines correspond to logarithmic fits, and $\sigma_O^N=\sigma_O/\overline{O}$. All panels: $\hat{H}_{\Delta=1,\lambda=0}$ and averages performed in the time interval $[100,500]$, except for the domain wall state which used $[5\times 10^2, 5\times 10^3]$.}
\label{fig:initial}	
\end{figure}

In Fig.~\ref{fig:initial}, as in Fig.~\ref{fig:shellH1}, the Hamiltonian is fixed rather than the initial state. We select the integrable isotropic $\hat{H}_{\Delta=1,\lambda=0}$. The value of $\kappa$ once again mirrors the width and filling of the energy shell (cf. Tables \ref{table:IPR}, \ref{table:kappaH1} and Figs.\ref{fig:shellH1}, \ref{fig:initial}). For the initial nonrandom states $|\text{DW}\rangle$, $|\text{NS}\rangle $, and $|\text{PP}\rangle $, the coefficient $\kappa$ is always much smaller than 0.35. None of these site-basis vectors behave as a chaotic state for $\hat{H}_{\Delta=1,\lambda=0}$.  The minimum $\kappa \sim 0.11$ occurs for $|\text{DW}\rangle$ and both observables, $\widehat{{\rm KE}}$ and $\hat{s}_f^z(\pi)$. Even here, the $R^2$ for the exponential fitting is slightly larger than for the power-law one.  Furthermore, the power-law fitting in this case has $b\sim 1.65$, which, again, is much larger than the 0.5 for the quasifree particle systems~\cite{Venuti2013,Cassidy2011,Gramsch2012,HeARXIVdisorder}.

The $|\text{PP}\rangle$ suffers from strong border effects. The first site of this state, always has a spin pointing down, but the spin on the last site can either point down, when $L/2$ is even, or up, when $L/2$ is odd. This causes the oscillations seen in Fig.~\ref{fig:initial} and the lower value of $R^2$ when compared to the other states.

\begin{table}[h]
\caption{Coefficient $\kappa$ in the exponential fittings $\sigma_O \propto e^{ - \kappa L}$ of Fig.~\ref{fig:initial}  for $\hat{H}_{\Delta=1,\lambda=0}$.}
\begin{center}
\begin{tabular}{|c|c|c|c|}
  \hline
  \multicolumn{4}{|c|}{$\kappa$ for $\hat{H}_{\Delta=1,\lambda=0}$} \\
\hline 
  $\hat{O}=$&  $\widehat{KE}$ &  $\hat{C}_{L/2,L/2+1}^z$ & $\hat{s}_f^z(\pi)$ \\
  \hline
$|\text{DW}\rangle $ & 0.109   & 0.133   & 0.109    \\
\hline
$|\text{NS}\rangle $ & 0.184   & 0.157  & 0.111     \\
\hline
$|\text{PP}\rangle $ &  0.246  & 0.215  &  0.244  \\
\hline
$|\xi_{{\cal S}^z = 0} \rangle$ & 0.354   &  0.331   & 0.325   \\
\hline
$|\xi_{2^L} \rangle$ &  0.370  &  0.343   & 0.345 \\
\hline
\end{tabular}
\end{center}
\label{table:kappaH1}
\end{table}

 Another feature that calls attention in Fig.~\ref{fig:initial} is the result for $|\text{DW}\rangle$ and $|\text{NS}\rangle$ for the structure factors: $\sigma_{s_f^{z(x)}}$ is larger for $|\text{DW}\rangle$ than for $|\text{NS}\rangle$. We notice, however, that $\kappa$ for the N\'eel state is larger, so the curves will eventually cross. This crossing is seen already for our system sizes for $\hat{H}_{\Delta=0.5,\lambda=1}$, for example (not shown).

\subsection{Exponential vs power-law decay}
\label{sec:compare}

\begin{figure}[hbt]
\centering
\includegraphics[width=0.45\textwidth]{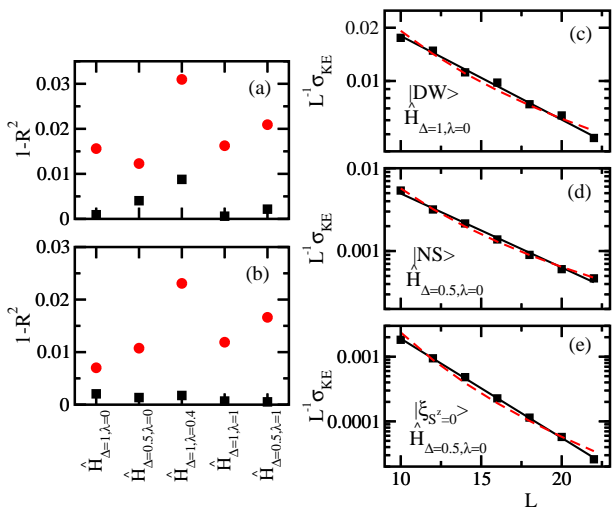}
\caption{(Color online)  Left panels: $1-R^2$ values for the (square) exponential and (circle) power-law fits of the decay of the temporal fluctuations of $\hat{C}^x_{L/2,L/2+1}$ with system size. The initial states are $|\text{NS}\rangle$ (a) and $|\xi_{{\cal S}^z = 0} \rangle$ (b).
Right panel: logarithmic plots of the standard deviation of the time fluctuations of $\widehat{{\rm KE}}$  vs $L$ for $|\text{DW}\rangle$ with $\hat{H}_{\Delta=1,\lambda=0}$ (c), $|\text{NS}\rangle$ with $\hat{H}_{\Delta=0.5,\lambda=0}$ (d), and $|\xi_{{\cal S}^z = 0} \rangle$ with $\hat{H}_{\Delta=0.5,\lambda=0}$ (e). The solid line corresponds to the fitting for the exponential decay and the dashed line corresponds to the power-law decay. The averages are performed in the time interval $[100,500]$, except for the domain wall state, which used $[5\times 10^2, 5\times 10^3]$.}
\label{fig:compare}	
\end{figure}

In Fig.~\ref{fig:compare}, we provide some examples for the comparison between the exponential and the power-law fitting. The left panels show the values of $1-R^2$ for the temporal fluctuations of the spin-spin correlation $\hat{C}^x_{L/2,L/2+1}$ for the five Hamiltonians considered. $1-R^2$ is at least 1 order of magnitude smaller for the exponential fitting and it reaches particularly small values when the initial state is thermal [Fig.~\ref{fig:compare}(b)].

The right panels show the decay of the fluctuations for $\widehat{{\rm KE}}$ with system size for three different combinations of initial states and Hamiltonians. The exponential fitting is visibly better, especially away from the pair domain wall and isotropic NN Hamiltonian. Also noticeable is the substantial decrease of $\sigma_{KE}$ for the same system size as we go from the top to the bottom panels, that is as the initial state further delocalizes.

\subsection{The XX model}

For the XX model, the bound in Eq.~(\ref{bound}) should not be valid anymore due to the many degeneracies of this model, as discussed in Sec.~\ref{sec:spectrum}. There are analytical and numerical studies supporting the power-law decay of the time fluctuations for systems of noninteracting particles~\cite{Venuti2013,Cassidy2011,Gramsch2012,HeARXIVdisorder}. In terms of numerics, since  quadratic Hamiltonians are trivially solvable, very large systems have been considered~\cite{Cassidy2011,Gramsch2012,HeARXIVdisorder}. Comparing the results of our numerical method with these previous findings  is, therefore, a good way to assess its validity.

For $|\text{DW}\rangle$ and $|\text{NS}\rangle$, the power-law fitting is, indeed, the best choice for some observables, but not all. More convincing here is the value of  $b$, which is more than twice as large for $\Delta \neq 0$ than for the XX model. For $\sum_{k} s_f^z(k)$, both states in fact lead to $b\sim 0.6$ for the noninteracting Hamiltonian, which is very close to the analytical prediction 0.5.

We emphasize that even for the XX Hamiltonian, the thermal initial states $|\xi_{{\cal S}^z = 0} \rangle$  and $|\xi_{2^L} \rangle$ clearly lead to exponential decays of the time fluctuations. This reinforces the importance of the initial state in studies of nonequilibrium dynamics, a point that has been explored more in the context of thermalization~\cite{Banuls2011,Fitzpatrick2011,Deng2011,HeInitial,TorresARXIV} and of the fluctuation-dissipation theorem~\cite{Khatami}  in isolated quantum systems.

Another difference between the noninteracting XX and the interacting XXZ model, which is concomitant to the differences in degeneracies, refers to the intrinsic nature of the fluctuations around the steady state. By analyzing the frequency spectrum of the fluctuations for the local magnetization on site $L/2$ using the fast Fourier transform, we see that the XX model has few well-defined narrow frequencies for both $|\text{DW}\rangle$ and $|\text{NS}\rangle$, which is to be contrasted to the XXZ model with $\Delta<1$. On the other hand, when the initial state is thermal, the spectrum is noisy, independent of the Hamiltonian.

\section{Relaxation} 
\label{sec:relaxation}

The smallest values of $\kappa$ for $\Delta \neq 0$ are associated with the domain wall. The coefficient decreases significantly as we go from $\Delta=0.5$ to $\Delta=1$, and then to $\Delta=1.5$ where the system is already in the gapped phase. The poor performance of this state reflects its proximity to the edge of the spectrum [Fig.~\ref{fig:shellH1} (a)] and consequent low connectivity. It is pushed there by the Ising interaction. The NN (NNN) Ising contribution to the energy of the site-basis vectors increases with the number of NN (NNN) pairs of parallel spins in the $z$ direction. The domain wall has the largest number of NN pairs, $L-2$, and it has $L-4$ NNN pairs [see Eq.~(\ref{Eq:DW})]. In terms of connectivity, the state is directly coupled to only one basis vector when $\hat{H}$ is the integrable Hamiltonian and only three vectors when $\lambda>0$. Thus, according to Eq.(\ref{deltaE}), the width of the shell is very small and it does not change with system size [see Eq.~(\ref{Eq:DWwidth})].

From the remarks above, its is clear that the relaxation process of  $|\text{DW}\rangle$ must be much slower than for the other initial states especially for large $\Delta$.  In fact, when $\Delta \gg 1 $, the domain wall freezes in time~\cite{DWstudies,Santos2011}. To study the temporal fluctuations, we needed to consider much longer time intervals than for the other initial states to guarantee that it had, indeed, relaxed. Moreover, the time taken to reach a steady state obviously increases with $L$, as we need to break more pairs of adjacent parallel spins~\cite{Santos2011}. 

\begin{figure}[htb]
\centering
\includegraphics[width=0.50\textwidth]{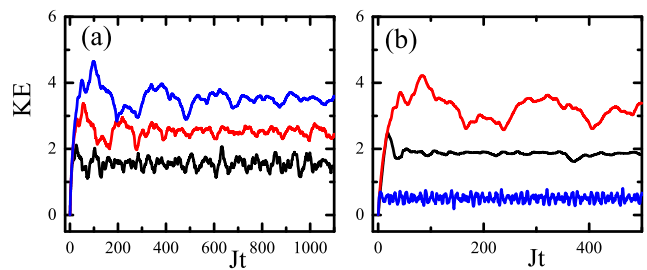}
\caption{(Color online) Time dependence of the kinetic energy for $| \text{ini} \rangle = |\text{DW}\rangle$. (a) $\hat{H}_{\Delta=1,\lambda=0}$, and $ L=24, 18, 12 $ from top to bottom.  (b) $\hat{H}_{\Delta = 1,\lambda=0}$, $\hat{H}_{\Delta = 0.5,\lambda=0}$, $\hat{H}_{\Delta = 1.5,\lambda=0}$ from top to bottom, $ L=22 $. }
\label{fig:DWrelax}	
\end{figure}

An illustration of the dependence on size and anisotropy is provided in Fig.~\ref{fig:DWrelax}. In panel (a), we see that the transient oscillations remain for longer times as $L$ increases. Here, a particularly bad combination is considered where the initial state is the domain wall and the Hamiltonian is  $\hat{H}_{\Delta=1,\lambda=0}$. When $\Delta=1$, the number of states taking part in the evolution is smaller than  for $\Delta \neq 0$, because, in addition to conservation of spin in the $z$-direction, there is also conservation of total spin. The special role of the isotropic point for $|\text{DW}\rangle$ is well illustrated in panel (b). For values away from the critical point, that is for $ \Delta =1.5 $ and $ \Delta =0.5 $, a steady state is reached after a few tens of $ Jt $, whereas for the case $ \Delta =1 $ the plateau is not reached even at  $ Jt\sim 500 $.

\begin{figure}[ht]
\centering
\includegraphics[width=0.42\textwidth]{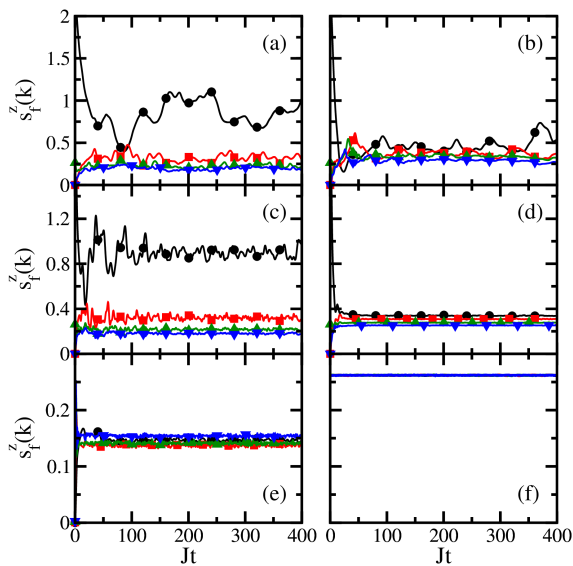}
\caption{(Color online) Relaxation of the structure factor in the $z$ direction; $L=22$, ${\cal S}^z=0$.
Momentum: (black circles) $k = 2\pi/11$, (red squares) $3\pi/11$, (green up triangles) $4\pi/11$, (blue down triangles)  $5\pi/11$. 
Top panels: $|\text{ini} \rangle = |\text{DW}\rangle$; (a) $\hat{H}_{\Delta=1,\lambda=0}$ and (b) $\hat{H}_{\Delta=0.5,\lambda=0}$. 
Middle panels: $|\text{ini} \rangle = |\text{DW}\rangle$; (c) $\hat{H}_{\Delta=1,\lambda=1}$ and (d) $\hat{H}_{\Delta=0.5,\lambda=1}$. 
Bottom panels: $\hat{H}_{\Delta=1,\lambda=0}$; (e) $| \text{ini} \rangle = |\text{NS}\rangle$ and (f) $|\xi_{{\cal S}^z = 0} \rangle$.}
\label{fig:relax}	
\end{figure}

In Fig.~\ref{fig:relax} we compare the relaxation process for $|\text{DW}\rangle$, $|\text{NS}\rangle$, and  $|\xi_{{\cal S}^z = 0} \rangle$. We choose the structure factor as observable because its time evolution has an interesting dependence on the momentum $k$ \cite{Canovinote}, depending on the initial state and the Hamiltonian evolving it.
In the top and middle panels, we show the evolution of  $\hat{s}_f^{z}(k)$ when  $| \text{ini} \rangle = |\text{DW}\rangle$. In the case of small $k$'s and at the isotropic point, the relaxation process is very slow and the fluctuations are large  [Figs.~\ref{fig:relax} (a) and (c)]. This occurs even for the strongly chaotic isotropic Hamiltonian $\hat{H}_{\Delta=1,\lambda=1}$ [Fig.~\ref{fig:relax} (c)]. If we break the symmetries associated with the isotropic point, the relaxation process becomes faster and the fluctuations become smaller for all $k$'s [Figs.~\ref{fig:relax} (b)], this being even better in the chaotic domain [Figs.~\ref{fig:relax} (d)]. Nevertheless, in all four panels, (a), (b), (c), and (d), the saturation value is not the same for all values of momentum, which suggests some residual memory of the initial state. 

Fast relaxation and small fluctuations occur for other initial states, even for the integrable isotropic Hamiltonian $\hat{H}_{\Delta=1,\lambda=0}$, provided  the width of the energy shell is not too narrow and $| \text{ini} \rangle$ is delocalized in the shell. This already is suggested by $|
\text{ini} \rangle = |\text{NS}\rangle$, although some reminiscent dependence on $k$ is still noticeable  [Fig.~\ref{fig:relax} (e)]. It becomes evident for the thermal state $|\xi_{{\cal S}^z = 0} \rangle$ [Fig.~\ref{fig:relax} (f)], where the dependence on $k$ is completely lost. Compare this behavior with the energy shells in Figs.~\ref{fig:shellH1} (b) and (d), respectively.

\section{Conclusions}    
Our results confirm that the exponential decay with $L$ of the time fluctuations of few-body observables after relaxation prevails in systems without excessive degeneracies, whether integrable or chaotic. The coefficient of this decay depends on the level of delocalization of the initial state with respect to the Hamiltonian dictating its evolution, in agreement with analytical predictions~\cite{Reimann2008,Short2011,Short2012,Reimann2012}. Therefore, it is not the initial state or the Hamiltonian separately that determines the size of the fluctuations, but the interplay between both. The quantification of this relation is embodied by the filling of the energy shell.

Interestingly, for the thermal initial states at infinite temperature, the exponential decay holds even for the noninteracting integrable model. 

Among the initial states considered, the domain wall has the smallest decay coefficient for the fluctuations and the slowest dynamics, especially when the system gets close to the isotropic point. This is a consequence of the presence of additional symmetries and the proximity of the state to the edge of the spectrum, where the density of states is low.  As $L$ increases, the domain wall takes longer to reach the steady state. The study of larger system sizes, which is essential to the absolute rejection of a power-law behavior for the time fluctuations, will be particularly challenging for this state. 

The initial states analyzed here can, in principle, be achieved in experiments with ultra cold atoms. The system sizes considered are also of relevance to these experiments, where tubes with as few as ten atoms are handled.

\begin{acknowledgments}
E.J.T.-H. and L.F.S. thank support by NSF via the grant  No.~DMR-1147430, as well as Felix Izrailev for discussions. A.I acknowledges support from CONICET (PIP 0662), ANPCyT (PICT 2010-1907) and UNLP (PID 11/X614), Argentina. P.R.Z., A.D.D. and H.M.P. acknowledge support from CONICET, ANPCyT, SeCyT-UNC, MinCyTCor and NVIDIA Professor Partnership Program.
\end{acknowledgments}


\begin{thebibliography}{56}
\bibitem[{pur()}]{pure}
\bibinfo{note}{We use the notation for a pure initial state, as considered in
  our numerics, but these results have been generalized also to mixed states.}
\end{thebibliography}
\end{document}